\documentclass[twocolumn,pra,showpacs,aps]{revtex4-1}
\usepackage{amssymb}
\usepackage{amsmath}
\usepackage{graphicx}
\usepackage{epsfig}

\begin{document}

\title{Projectively topological exceptional points in non-Hermitian
Rice-Mele model}
\author{C. Li${}^2$ and Z. Song${}^1$}
\email{songtc@nankai.edu.cn}
\affiliation{${}^1$School of Physics, Nankai University, Tianjin 300071, China \\
${}^2$Department of Physics, University of Hong Kong, Hong Kong, China}

\begin{abstract}
We study coupled non-Hermitian Rice-Mele chains, which consist of
Su-Schrieffer-Heeger (SSH) chain system with staggered on-site imaginary
potentials. In two dimensional ($2$D) thermodynamic limit, the exceptional
points (EPs) are shown to exhibit topological feature: EPs correspond to
topological defects of a real auxiliary $2$D vector field in $\boldsymbol{k}$
space, which is obtained from the Bloch states of the non-Hermitian
Hamiltonian. As a topological invariant, the topological charges of EPs can
be $\pm 1/2$, obtained by the winding number calculation. Remarkably, we
find that such a topological characterization remains for a finite number of
coupled chains, even a single chain, in which the momentum in one direction
is discrete. It shows that the EPs in the quasi-$1$D system still exhibit
topological characteristics and can be an abridged version for a $2$D system
with symmetry protected EPs that are robust in perturbations, which proves
that topological invariants for a quasi-$1$D system can be extracted from
the projection of the corresponding $2$D limit system on it.
\end{abstract}

\maketitle




\section{Introduction}

Exceptional point (EP) is an exclusive critical point in non-Hermitian
systems, at which many exotic features occur \cite%
{Bender,Longhi,West,LC,Yang,Zhang,Longhi1,Goldzak,Yi}. It is also called
non-Hermitian degeneracy or branch point in the complex energy plane, where
two eigenvalues and their eigenvectors become the same, referred to as a
coalescing level \cite{Heiss}. Mathematically, the occurrence of an EP\
relates to the emergence of a Jordan block. Two eigenvectors of a $2\times 2$
Jordan block are the same and self-orthogonal \cite{Ali1}. Unlike the level
crossing in the Hermitian matrix around the degenerate point, there is a
level repulsion near the EP. In general, EPs are sensitive to the parameters
of the system. In a continuous system with translational symmetry, the eigen
problem is reduced to that of a small non-Hermitian system, where momentum $%
k $ acts as a system parameter \cite{Ali2}.

With these unusual properties, recently there have been growing efforts to
investigate topological phenomena of non-Hermitian systems, both
theoretically \cite{Esaki,Tony,Li1,Lin,Nori,Fu,Yao,Gong} and experimentally
\cite{Ding,Dop,Weim,Feng}. Compared with Hermitian systems, of which
degenerate points play an important role in topological properties \cite%
{Niu,Kane,ZhangSQ,Burkov,Xu,Young,Sama,Weng,Huang,Hou,Liu,Neupane,SXu,Lv,Lu}%
, no effective method such as the calculation of the winding number or Chern
number in Hermitian systems \cite{Niu,Kane,ZhangSQ} is established to
describe the topological feature of EPs. Based on this, Almost all
theoretical works try to define or find a group of new topological invariant
to distinguish the phase diagram of non-Hermitian systems, some of them are
applicable for families of non-Hermitian Hamiltonians \cite%
{Esaki,Nori,Fu,Gong}, others focus on the specific model \cite%
{Tony,Li1,Lin,Yao}. Therefore, the description of the topological feature of
EPs remains an open question.

In this work, we study $M$ coupled non-Hermitian Rice-Mele chains with
length $2N$ \cite{RM,Lin1,WR}. The non-Hermiticity arises from staggered
on-site imaginary potentials on the Hermitian coupled Su-Schrieffer-Heeger
(SSH) chains \cite{SSH,As}. The exact solution shows that the EPs can be two
isolated points in two dimensional ($2$D) $\boldsymbol{k}$-plane for
infinite $M$ and $N$, or $2$D thermodynamic limit. When parameters of the
system vary, such two points move in the plane and cannot be removed until
they meet together. We map the Bloch states of the non-Hermitian Hamiltonian
onto a $2$D real vector field in $\boldsymbol{k}$-plane, referred to as an
auxiliary field. It is shown that two isolated EPs are topological defects
of the field with a topological charge $\pm 1/2$, which exhibits the
topological feature of the EPs. Furthermore, we extend this analysis to
finite $M$ cases with periodic and open boundary conditions. The
distribution of the auxiliary field at a pair of EPs in the systems with
even $M$ and periodic boundary condition (or odd $M$\ with open boundary
condition) reflects the same topological configuration, which indicates that
the auxiliary field for finite $M$ on the discrete $\boldsymbol{k}$ space is
the projection of the infinite one for the $2$D model. In this sense, the
topological charge obtained from the $2$D model\ in the thermodynamic limit
can characterize the EPs for finite $M$, even for $M=1$. Therefore, we focus
on the single-chain system and find that two topological defects in $2$D $%
\boldsymbol{k}$ space reduce to a pair of kinks in $1$D $k$ space. The
corresponding topological invariant is protected by the combined inversion
and time-reversal symmetry. At last, the behavior of the topological gapless
system in the presence of several types of perturbations is performed, which
shows the robust topological feature of the EPs in non-Hermitian systems is
the same as the band touching points in Hermitian systems.

The remainder of this paper is organized as follows. In Sec. \ref{Model and
phase diagram}, we present $M$ coupled non-Hermitian Rice-Mele chains with
length $2N$ and its phase diagram. Sec. \ref{Topological nodal points}
reveals the topological feature of nodal points of the $2$D system in the
thermodynamic limit and corresponding topological configuration in the
finite $2$D system. Sec. \ref{Single-chain situation} shows the single-chain
solution, including the energy band, eigenvector, and topological
properties. Sec. \ref{Symmetry protection of kinks} devotes to the
symmetries that protect the topological invariants of the single SSH chain.
Sec. \ref{Perturbations} displays the behavior of the topological gapless
system in the presence of several types of perturbations. Finally, we
present a summary and discussion in Sec. \ref{Summary}.

\section{Model and Phase diagram}

\label{Model and phase diagram}

\begin{figure}[tbp]
\includegraphics[ bb=83 461 424 743, width=0.45\textwidth, clip]{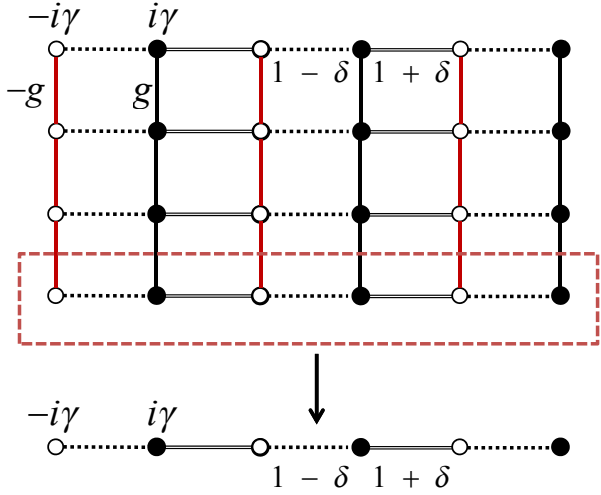}
\caption{(Color online) Schematics of the simplest bundle of $M$
non-Hermitian dimerized chains with staggered balanced gain and loss.
Hopping amplitudes along the $x$ direction are staggered by $1+\protect%
\delta $ (double line) and $1-\protect\delta $ (dash line). Along the $y$
direction, it is a uniform chain with inter-chain tunneling constant $g$ ($%
-g $). When the system reduces into the $1$D situation, it can be seen as a
SSH chain with on-site imaginary potentials, which will be explored in the
rest of the paper.}
\label{fig1}
\end{figure}

\begin{figure}[tbp]
\includegraphics[ bb=38 332 466 750, width=0.45\textwidth, clip]{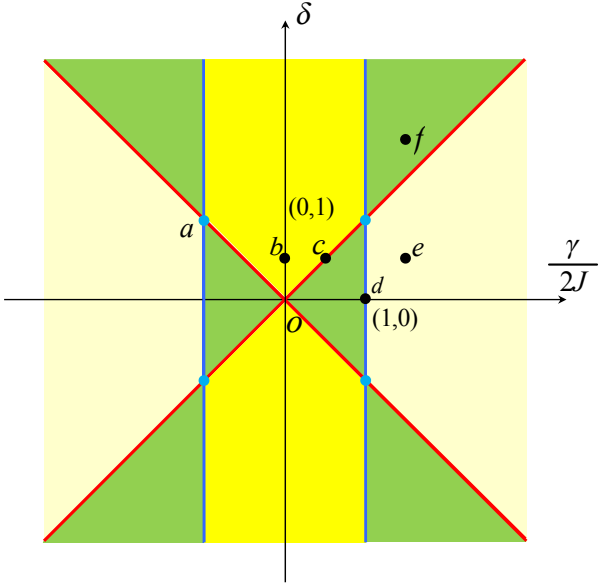}
\caption{(Color online) Phase diagram of coupled SSH chain system on the $%
\protect\delta -\protect\gamma /2J$ plane (in the unit of $J$) when $V_{%
\mathbf{k}}=0$. Red and blue lines indicate the boundary represented in Eq. (%
\protect\ref{boundary}), which separated the area with EPs exist (green), $%
\protect\varepsilon _{k}$ is real (yellow), and $\protect\varepsilon _{k}$
is imagined (ashen). Four blue circles mean the touching points of two
different boundaries.}
\label{fig2}
\end{figure}

We consider a bundle of $M$ non-Hermitian dimerized chains with staggered
balanced gain and loss, two neighboring of which are coupled. The simplest
tight-binding model with these features is%
\begin{eqnarray}
H &=&-J\sum_{l=1}^{2N}\sum_{n=1}^{M}\left[ 1+\left( -1\right) ^{l}\delta %
\right] \left( c_{l,n}^{\dagger }c_{l+1,n}+\mathrm{H.c.}\right)  \notag \\
&&+g\sum_{l=1}^{2N}\sum_{n=1}^{M}\left( -1\right) ^{l}\left(
c_{l,n}^{\dagger }c_{l,n+1}+\mathrm{H.c.}\right)  \notag \\
&&+i\gamma \sum_{l,n}\left( -1\right) ^{l}c_{l,n}^{\dagger }c_{l,n},
\end{eqnarray}%
where $J$, $\delta $, $g$ and $i\gamma $ ($\gamma >0$), as showed in Fig. %
\ref{fig1}, are the inter-chain hopping strengths, the distortion factor,
inter-chain tunneling constant and the alternating imaginary potential
magnitude, respectively. Here $c_{l,n}^{\dagger }$ is the creation operator
of the fermion at the $l$th site in $n$th chain. The periodic boundary
conditions along two directions are imposed as $c_{2N+1,n}=c_{1,n}$ and $%
c_{l,M+1}=c_{l,1}$. We decompose the system into two sub-lattices $A$ and $B$
and rewrite the Hamiltonian as%
\begin{eqnarray}
H &=&-J\sum_{l=1}^{N}\sum_{n=1}^{M}\left[ \left( 1-\delta \right)
a_{l,n}^{\dagger }b_{l,n}+\left( 1+\delta \right) b_{l,n}^{\dagger
}a_{l+1,n}+\mathrm{H.c.}\right]  \notag \\
&&-g\sum_{l=1}^{N}\sum_{n=1}^{M}\left( a_{l,n}^{\dagger
}a_{l,n+1}-b_{l,n}^{\dagger }b_{l,n+1}+\mathrm{H.c.}\right)  \notag \\
&&+i\gamma \sum_{l,n}\left( b_{l,n}^{\dagger }b_{l,n}-a_{l,n}^{\dagger
}a_{l,n}\right) ,
\end{eqnarray}%
where $a_{l}^{\dag }$ and $b_{l}^{\dag }$ are the creation operators of
fermion at $l$th site of sub-lattice $A$ and $B$, respectively. Taking the
Fourier transformations

\begin{equation}
\left\{
\begin{array}{c}
a_{\boldsymbol{k}}=\frac{1}{\sqrt{NM}}\sum_{l,n}e^{i(k_{x}l+k_{y}n)}a_{l,n}
\\
b_{\boldsymbol{k}}=\frac{1}{\sqrt{NM}}\sum_{l,n}e^{i(k_{x}l+k_{y}n)}b_{l,n}%
\end{array}%
\right. ,
\end{equation}%
where $\boldsymbol{k}=(k_{x},k_{y})$ $=(2\pi n_{x}/N-\pi ,2\pi n_{y}/M-\pi )$%
, with $n_{x}=0,1,2,\ldots ,N-1,n_{y}=0,1,2,\ldots ,M-1$, we have
\begin{equation}
H=\sum_{\boldsymbol{k}}H_{\boldsymbol{k}}=\sum_{\boldsymbol{k}}(a_{%
\boldsymbol{k}}^{\dagger },b_{\boldsymbol{k}}^{\dagger })h_{\boldsymbol{k}%
}\left(
\begin{array}{c}
a_{\boldsymbol{k}} \\
b_{\boldsymbol{k}}%
\end{array}%
\right) .
\end{equation}%
Here we just let $J=1$ for simplicity. The core matrix is%
\begin{equation}
h_{\boldsymbol{k}}=-\left(
\begin{array}{cc}
V_{k}+i\gamma & \left( 1-\delta \right) +\left( 1+\delta \right) e^{-ik_{x}}
\\
\left( 1-\delta \right) +\left( 1+\delta \right) e^{ik_{x}} & -\left(
V_{k}+i\gamma \right)%
\end{array}%
\right) ,  \label{core}
\end{equation}%
where%
\begin{equation}
V_{\boldsymbol{k}}=2g\cos k_{y}.
\end{equation}%
The spectrum is%
\begin{equation}
\varepsilon _{k}=\pm 2\sqrt{\left( 1-\delta ^{2}\right) \cos ^{2}\left(
\frac{k_{x}}{2}\right) +\delta ^{2}+\left( g\cos k_{y}+\frac{i\gamma }{2}%
\right) ^{2}}
\end{equation}%
with the eigenvector%
\begin{equation}
\left\vert \psi _{k}^{\pm }\right\rangle =\frac{1}{\Omega _{\pm }}\left(
\begin{array}{c}
i\gamma +V_{k}\mp \varepsilon _{k} \\
\left( 1-\delta \right) +\left( 1+\delta \right) e^{ik_{x}}%
\end{array}%
\right) ,
\end{equation}%
which are normalized by the Dirac normalization factor%
\begin{eqnarray}
\Omega _{\pm } &=&\left\vert \varepsilon _{k}\right\vert ^{2}\mp \left(
2\gamma \mathrm{Im}\varepsilon _{k}+2V_{\mathbf{k}}\mathrm{Re}\varepsilon
_{k}\right) +V_{k}^{2}+\gamma ^{2}  \notag \\
&&+\varepsilon _{k}^{2}-\left( V_{k}+i\gamma \right) ^{2}.
\end{eqnarray}%
From $\varepsilon _{k}=0$, we have equations

\begin{equation}
\left\{
\begin{array}{c}
\left( 1-\delta ^{2}\right) \cos ^{2}\left( \frac{k_{x}}{2}\right) +\delta
^{2}-\left( \frac{\gamma }{2}\right) ^{2}=0 \\
2g\cos k_{y}=0%
\end{array}%
\right. .
\end{equation}%
In $\boldsymbol{k}$\ space the zero energy points are located at $%
\boldsymbol{k}_{c}=(k_{cx},k_{cy})$:%
\begin{equation}
\left\{
\begin{array}{c}
\frac{\left( \frac{\gamma }{2}\right) ^{2}-\delta ^{2}}{1-\delta ^{2}}=\cos
^{2}\left( \frac{k_{cx}}{2}\right) \\
k_{cy}=\pm \pi /2%
\end{array}%
\right. .
\end{equation}%
The restriction $\left\vert \cos k_{cx}\right\vert \leqslant 1$\ leads to%
\begin{equation}
(\gamma ^{2}-4\delta ^{2})(\gamma ^{2}-4)\leqslant 0,
\end{equation}%
then the boundaries are%
\begin{equation}
\left\{
\begin{array}{c}
\gamma _{c}=\pm 2\delta \\
\gamma _{c}=\pm 2%
\end{array}%
\right. .  \label{boundary}
\end{equation}%
Here we note that the boundary line is independent of $g$, so we set $g=1$
for convenience in the rest of the paper. It can be found that the above
boundaries divide the parameter space into three parts when $V_{k}=0$: the
broken area (where EPs appear), the real and imagine eigenvalues area (where
always have an energy gap), which is displayed in Fig. \ref{fig2}. We will
show it more clearly in the discussion of single-chain model below.

\section{Topological nodal points}

\label{Topological nodal points}

In this section, we will show that the EPs have topological properties. We
demonstrate this point by rewriting the core matrix $h_{\boldsymbol{k}}$
from Eq. (\ref{core}) in the form

\begin{equation}
h_{\boldsymbol{k}}=\mathbf{B}\left( \boldsymbol{k}\right) \cdot \mathbf{%
\sigma }_{\boldsymbol{k}},  \label{H_BS}
\end{equation}%
where the components of the auxiliary field $\mathbf{B}\left( \boldsymbol{k}%
\right) =(B_{x},B_{y},B_{z})$\ are%
\begin{equation}
\left\{
\begin{array}{l}
B_{x}=-\left[ \left( 1-\delta \right) +\left( 1+\delta \right) \cos k_{x}%
\right] \\
B_{y}=-\left( 1+\delta \right) \sin k_{x} \\
B_{z}=-(V_{k}+i\gamma )%
\end{array}%
\right. .
\end{equation}%
The Pauli matrices $\mathbf{\sigma }_{\boldsymbol{k}}$\ are taken as the form

\begin{equation}
\sigma _{x}=\left(
\begin{array}{cc}
0 & 1 \\
1 & 0%
\end{array}%
\right) ,\sigma _{y}=\left(
\begin{array}{cc}
0 & -i \\
i & 0%
\end{array}%
\right) ,\sigma _{z}=\left(
\begin{array}{cc}
1 & 0 \\
0 & -1%
\end{array}%
\right) ,
\end{equation}%
and $B_{x}$ and $B_{y}$ are real. At the zero energy point, we note that
\begin{equation}
\varepsilon _{k}=\left\langle h_{\boldsymbol{k}}\right\rangle _{\boldsymbol{k%
}}=B_{x}\left\langle \sigma _{x}\right\rangle _{\boldsymbol{k}%
}+B_{y}\left\langle \sigma _{y}\right\rangle _{\boldsymbol{k}%
}+B_{z}\left\langle \sigma _{z}\right\rangle _{\boldsymbol{k}}=0,
\end{equation}%
and it requires
\begin{equation}
B_{z}=0\text{ or }\left\langle \sigma _{z}\right\rangle _{\boldsymbol{k}}=0,
\end{equation}%
where $\left\langle \sigma _{\alpha }\right\rangle _{\boldsymbol{k}%
}=\left\langle \psi _{\boldsymbol{k}}^{-}\right\vert \sigma _{\alpha
}\left\vert \psi _{\boldsymbol{k}}^{-}\right\rangle $ is the expectation
value of $\sigma _{\alpha }$\ $(\alpha =x,y,z)$. EPs occur when%
\begin{equation}
B_{x}\left\langle \sigma _{x}\right\rangle _{\boldsymbol{k}%
}+B_{y}\left\langle \sigma _{y}\right\rangle _{\boldsymbol{k}%
}=0,\left\langle \sigma _{z}\right\rangle _{\boldsymbol{k}}=0.
\end{equation}%
We introduce a $2$D real vector field $\mathbf{F}(\boldsymbol{k})$\ in $%
\boldsymbol{k}$ space, which is defined as

\begin{equation}
\mathbf{F}(\boldsymbol{k})=(B_{x}\left\langle \sigma _{x}\right\rangle _{%
\boldsymbol{k}}+B_{y}\left\langle \sigma _{y}\right\rangle _{\boldsymbol{k}%
},\left\langle \sigma _{z}\right\rangle _{\boldsymbol{k}}).  \label{field}
\end{equation}%
The components of the field can be directly obtained%
\begin{equation}
\left\{
\begin{array}{l}
F_{x}=-\frac{2\left[ \varepsilon _{k}^{2}-\left( V_{\mathbf{k}}+i\gamma
\right) ^{2}\right] \left( \mathrm{Re}\varepsilon _{k}+V_{k}\right) }{%
\varepsilon _{k}^{2}+\left\vert \varepsilon _{k}\right\vert ^{2}+2\gamma
\mathrm{Im}\varepsilon _{k}+2V_{k}\mathrm{Re}\varepsilon _{k}+2\gamma
^{2}-2i\gamma V_{k}} \\
F_{y}=\frac{\left\vert \varepsilon _{k}\right\vert ^{2}-\varepsilon
_{k}^{2}+2\gamma \mathrm{Im}\varepsilon _{k}+2V_{k}\mathrm{Re}\varepsilon
_{k}+2i\gamma V_{k}+2V_{k}^{2}}{\varepsilon _{k}^{2}+\left\vert \varepsilon
_{k}\right\vert ^{2}+2\gamma \mathrm{Im}\varepsilon _{k}+2V_{k}\mathrm{Re}%
\varepsilon _{k}+2\gamma ^{2}-2i\gamma V_{k}}%
\end{array}%
\right. .
\end{equation}

\begin{figure*}[tbp]
\includegraphics[ bb=54 297 461 650, width=0.3\textwidth, clip]{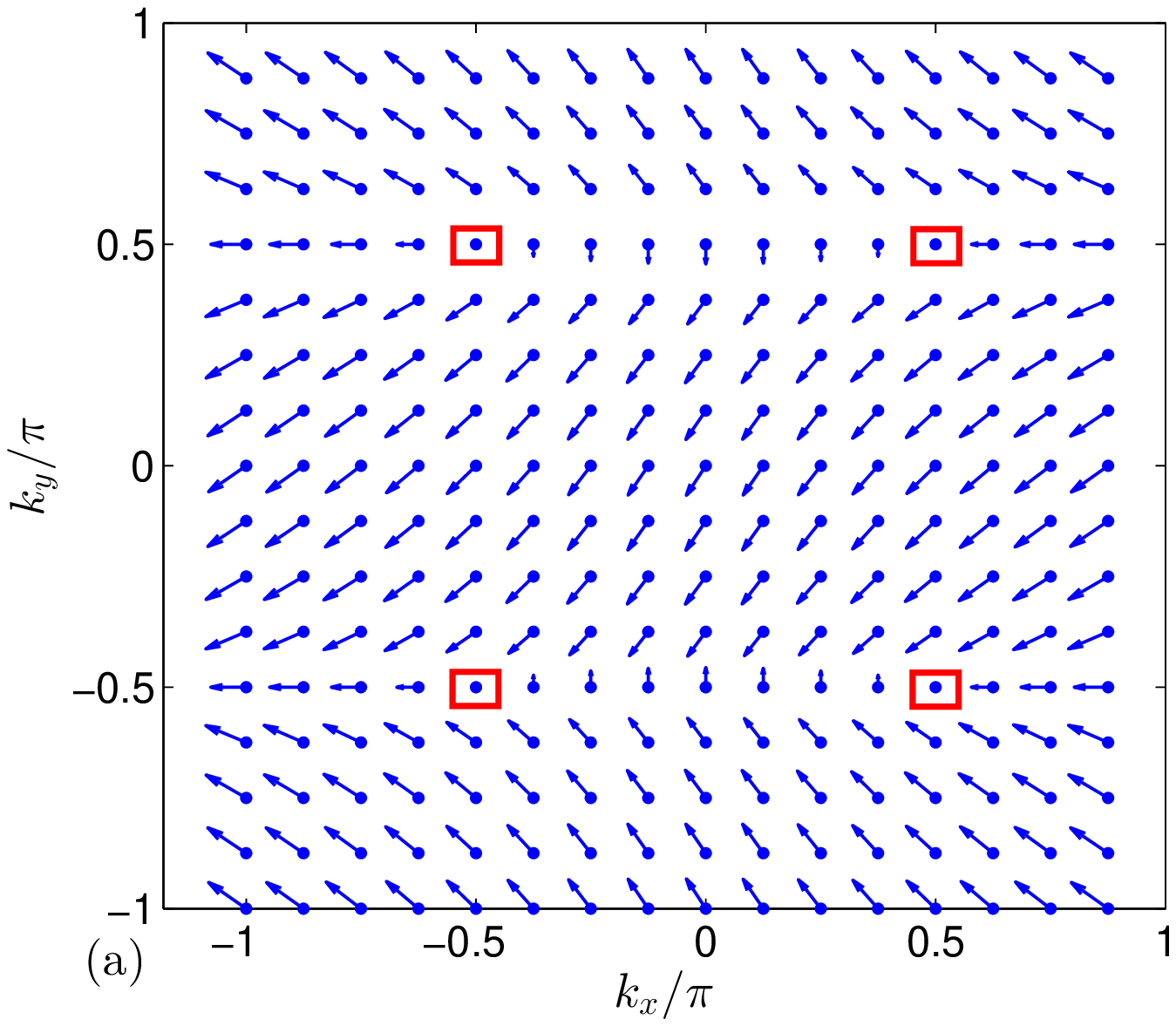} %
\includegraphics[ bb=54 297 461 650, width=0.3\textwidth, clip]{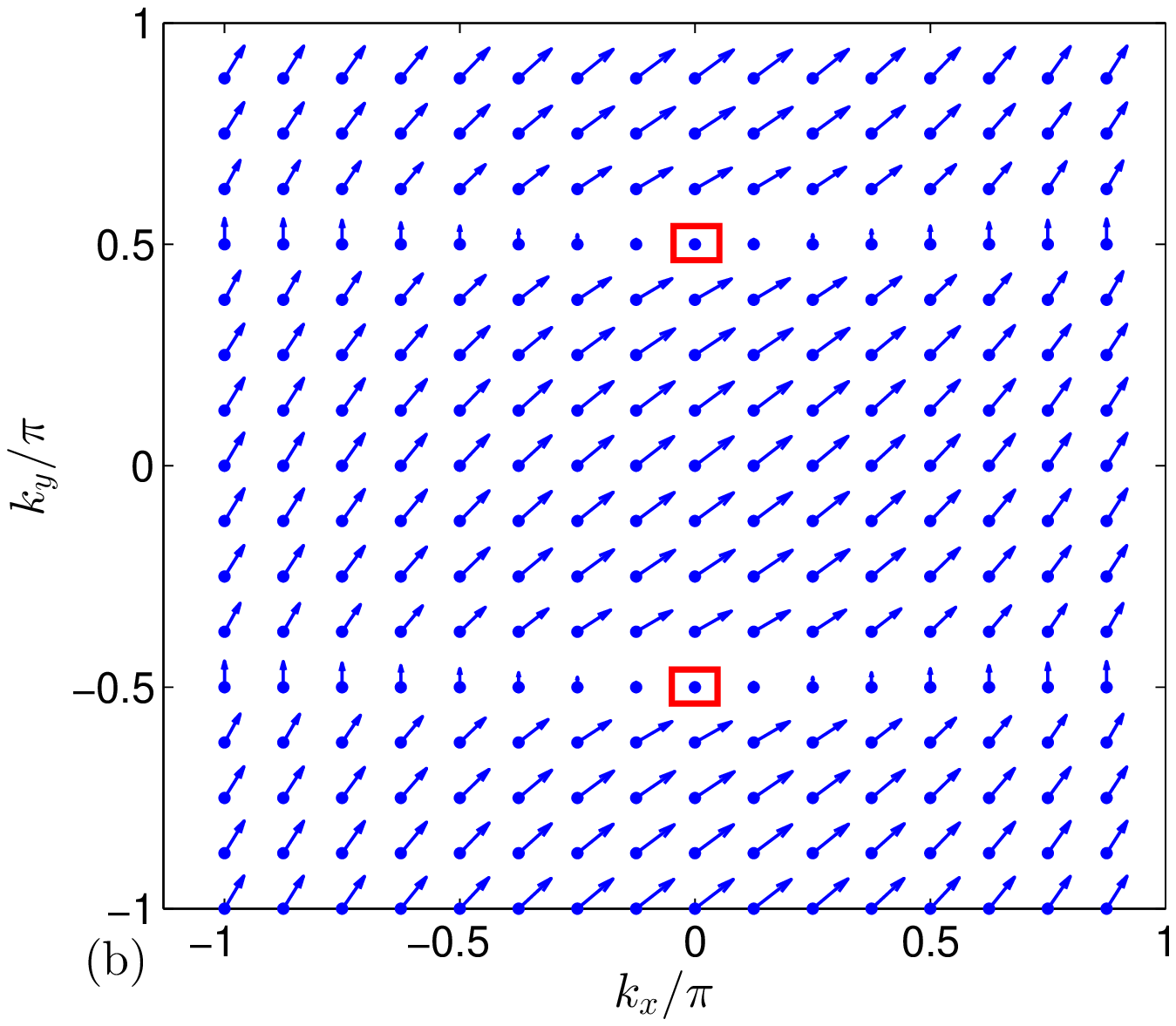} %
\includegraphics[ bb=54 297 461 650, width=0.3\textwidth, clip]{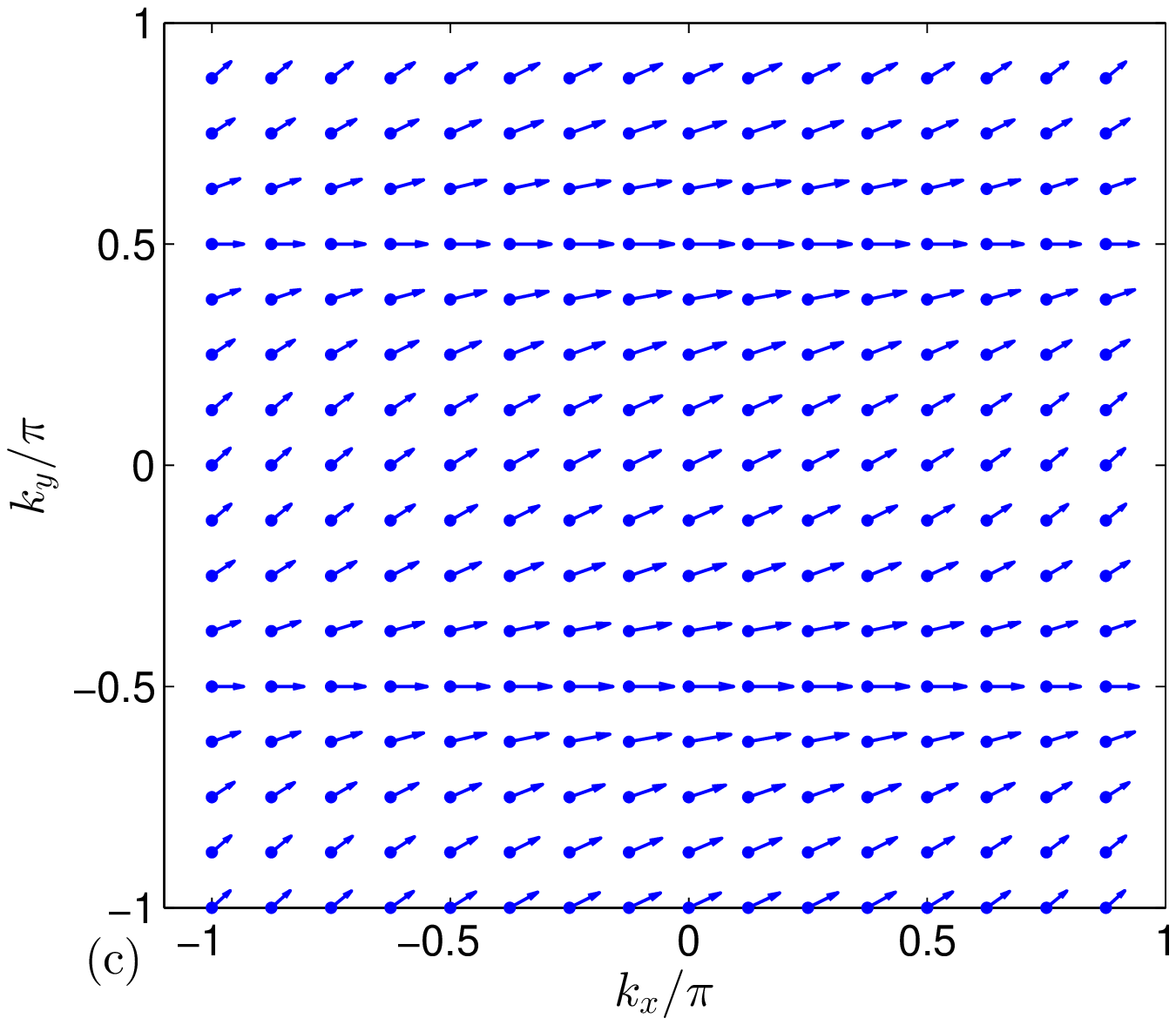}
\caption{The vortex structure of the EPs in $\boldsymbol{k}$ plane,
described by the field distribution based on the Eq. (\protect\ref{field})
with $N=M=16$. The parameter choices are (a)$\protect\gamma =\protect\sqrt{3}%
,\protect\delta =1/\protect\sqrt{2}$ in the area EPs exist; (b)$\protect%
\gamma =2,\protect\delta =1/\protect\sqrt{2}$ in the phase boundary; (c)$%
\protect\gamma =0,\protect\delta =1/\protect\sqrt{2}$ in the unbroken real
energy area, with the unit of $J$. The red square in (a) and (b) show the
position of EPs. Those figures correspond with the three situations we
mentioned before.}
\label{fig3}
\end{figure*}

In condensed matter physics, the Dirac or Weyl point acts like a singularity
of the Berry curvature in the Brillouin zone, or a magnetic monopole in $%
\boldsymbol{k}$ space. When a degenerate point is isolated, it should be a
vortex of the Berry curvature as a vector field, which is the topological
defect of the field \cite{Kane1,Zhou}. In parallel, when $M$ and $N$ are
infinite or reach the $2$D limit, the appearance of an EP in the present
model can be regarded as a field defect. The topological invariant of a
defect is the winding number

\begin{equation}
w=\frac{1}{2\pi }\oint_{C}d\boldsymbol{k}\mathbf{(}\hat{F}_{y}\nabla \hat{F}%
_{x}-\hat{F}_{x}\nabla \hat{F}_{y}),
\end{equation}%
where the unit vector $\mathbf{\hat{F}}(\boldsymbol{k})=\mathbf{F}\left(
\boldsymbol{k}\right) /\left\vert \mathbf{F}\left( \boldsymbol{k}\right)
\right\vert $ and $\nabla =\partial /\partial \boldsymbol{k}$ is the nabla
operator in $\boldsymbol{k}$ space. It is easy to check that when the
integral loop does not across the EPs, $w$\ always equals zero. Once the
loop across the EPs, we can always get an approximate expression of $\left(
F_{x},F_{y}\right) $\ when it is close to the EPs, where%
\begin{eqnarray}
F_{x} &\approx &2\gamma \mathrm{Re}e,F_{y}\approx 2\gamma \mathrm{Im}e,
\notag \\
e &=&\sqrt{4i\gamma y\sin k_{cy}-2vwx\sin k_{cx}}.
\end{eqnarray}%
Here $\left( k_{cx},k_{cy}\right) $\ is the coordinate of the EP in the
momentum space, $y=k_{y}-k_{cy},x=k_{x}-k_{cx}$. Based on this
approximation, the straightforward calculation tells us that when the loop
across the EPs,%
\begin{equation}
w=\pm \frac{1}{2}.
\end{equation}%
The same topological feature can be reflected in the finite system. We show
this point by the vortex structure of the EPs in $k$-plane in Fig. \ref{fig3}%
. Two types of topological configurations appear, corresponding to the
broken area and the boundary $\gamma _{c}=2$ in Fig. \ref{fig2}, separately.
One of them is two pairs of vortices with opposite chirality, the other is
one pair of vortices with the same chirality, as shown in Fig. \ref{fig3}(a)
and (b). The unbroken areas when $V_{k}=0$ is a trivial case that is no
vortex, as shown in Fig. \ref{fig3}(c).

Also, if we consider the open boundary condition in the $y$ direction when $%
M $ is odd and keep other conditions as same as before, the only change
would be $k_{y}\rightarrow \frac{\pi n}{M+1}.$For $n=(M+1)/2,$\ there are
still two EPs with topological features, as the projection from the $2$D
thermodynamic limit.

\section{Single-chain situation}

\label{Single-chain situation}

The above analysis is for the $2$D system. However, we note that the
position of EPs is only restricted to $k_{cy}=\pm \pi /2$. A straightforward
derivation is that the topological feature of EP in the $2$D system can be
retrieved from coupled-chain or quasi-$1$D system. Based on this idea, we
consider $M=1$, i.e., a single-chain Hamiltonian below
\begin{eqnarray}
H &=&-J\sum_{l=1}^{2N}\left[ 1+\left( -1\right) ^{l}\delta \right] \left(
c_{l}^{\dagger }c_{l+1}+\mathrm{H.c.}\right)  \label{Ham0} \\
&&+i\gamma \sum_{l}\left( -1\right) ^{l}c_{l}^{\dagger }c_{l},  \notag
\end{eqnarray}%
with the periodic boundary condition. Here, without loss of generality, we
define the action of time-reversal and parity in such a ring system as
follows. While the time-reversal operation $\mathcal{T}$ is defined as $%
\mathcal{T}i\mathcal{T}=-i$, the effect of the parity is $\mathcal{P}%
c_{l}^{\dagger }\mathcal{P}=c_{2N+1-l}^{\dagger }$.

\begin{figure*}[tbp]
\includegraphics[ bb=25 242 340 511, width=0.3\textwidth, clip]{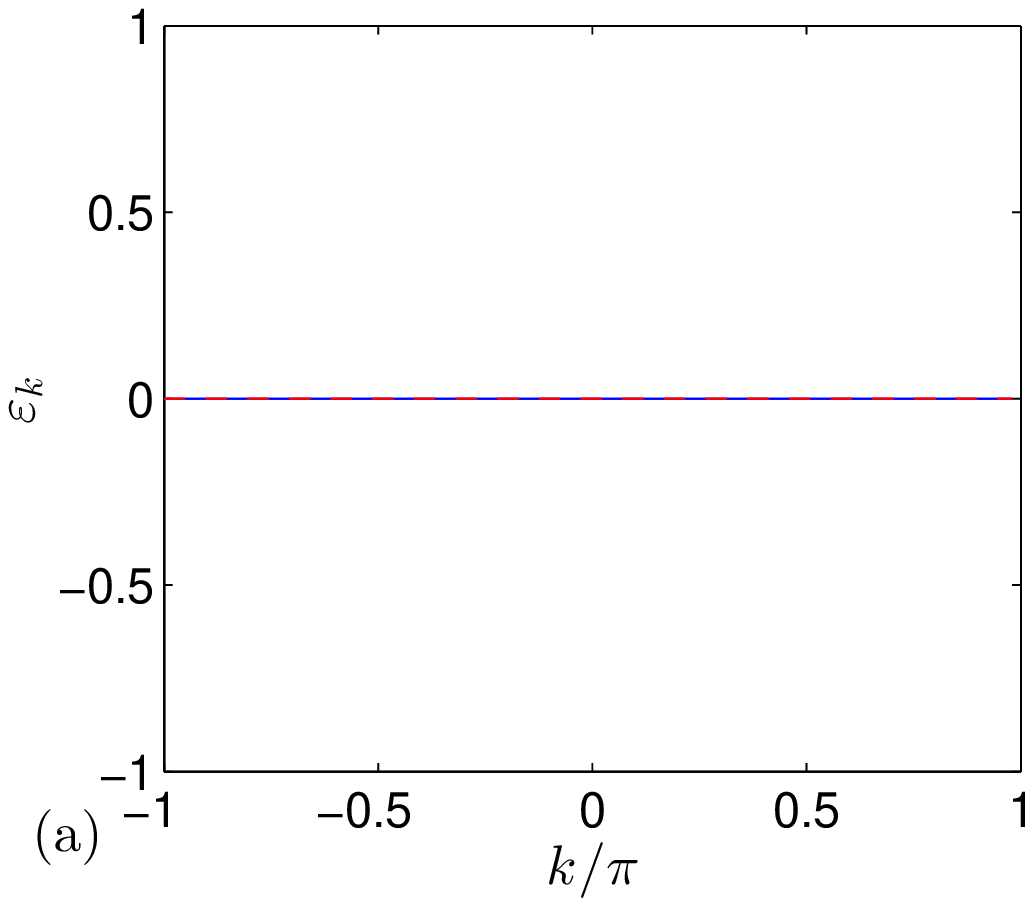} %
\includegraphics[ bb=25 242 340 511, width=0.3\textwidth, clip]{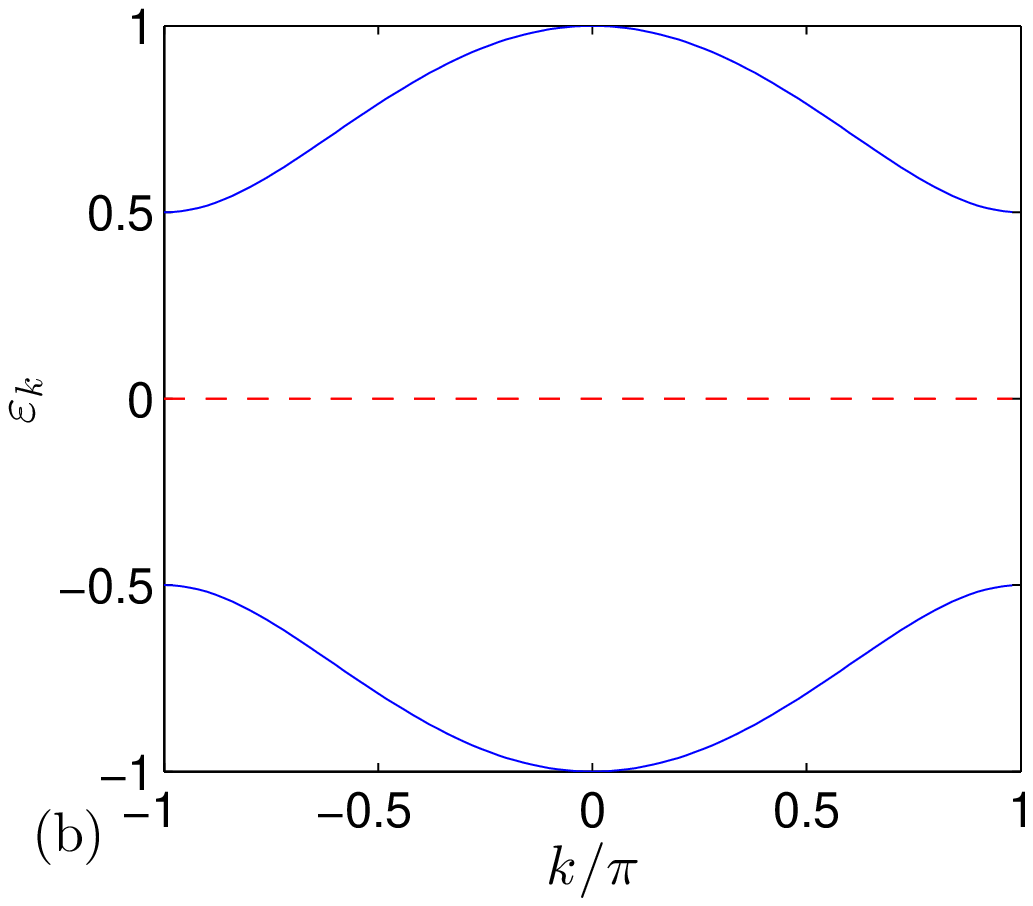} %
\includegraphics[ bb=25 242 340 511, width=0.3\textwidth, clip]{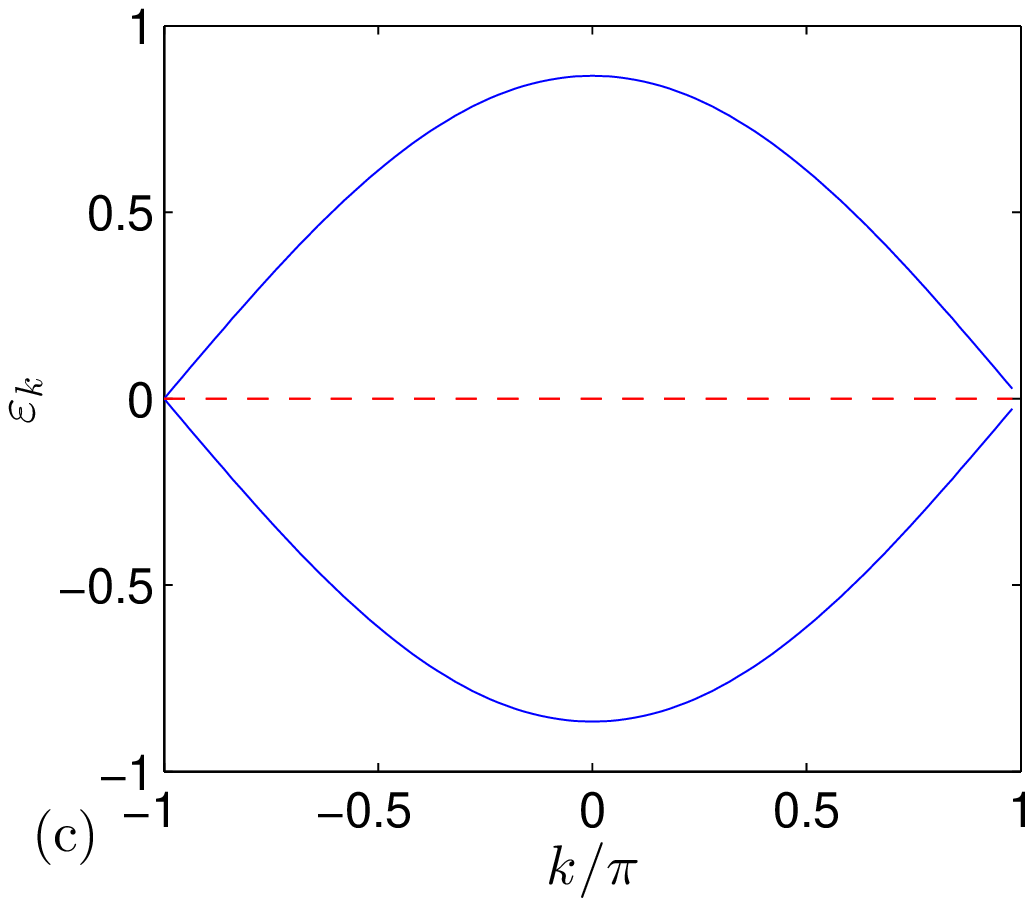} %
\includegraphics[ bb=25 242 340 511, width=0.3\textwidth, clip]{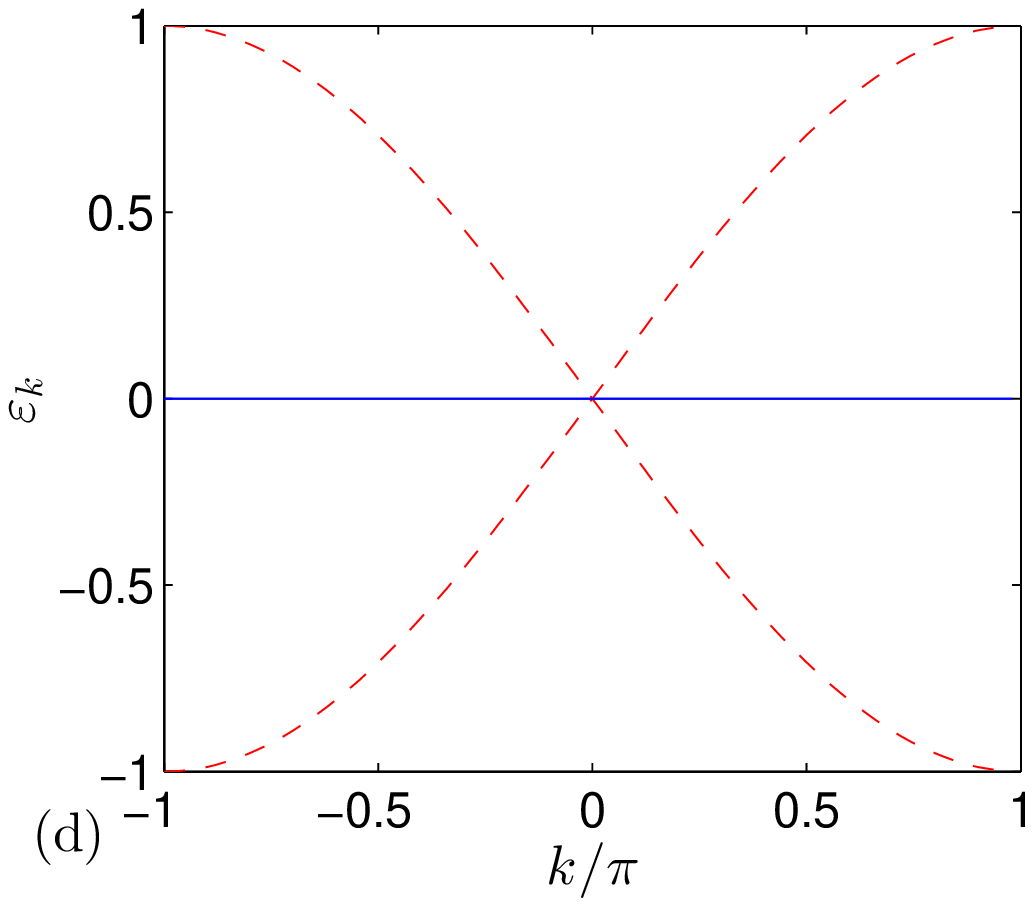} %
\includegraphics[ bb=25 242 340 511, width=0.3\textwidth, clip]{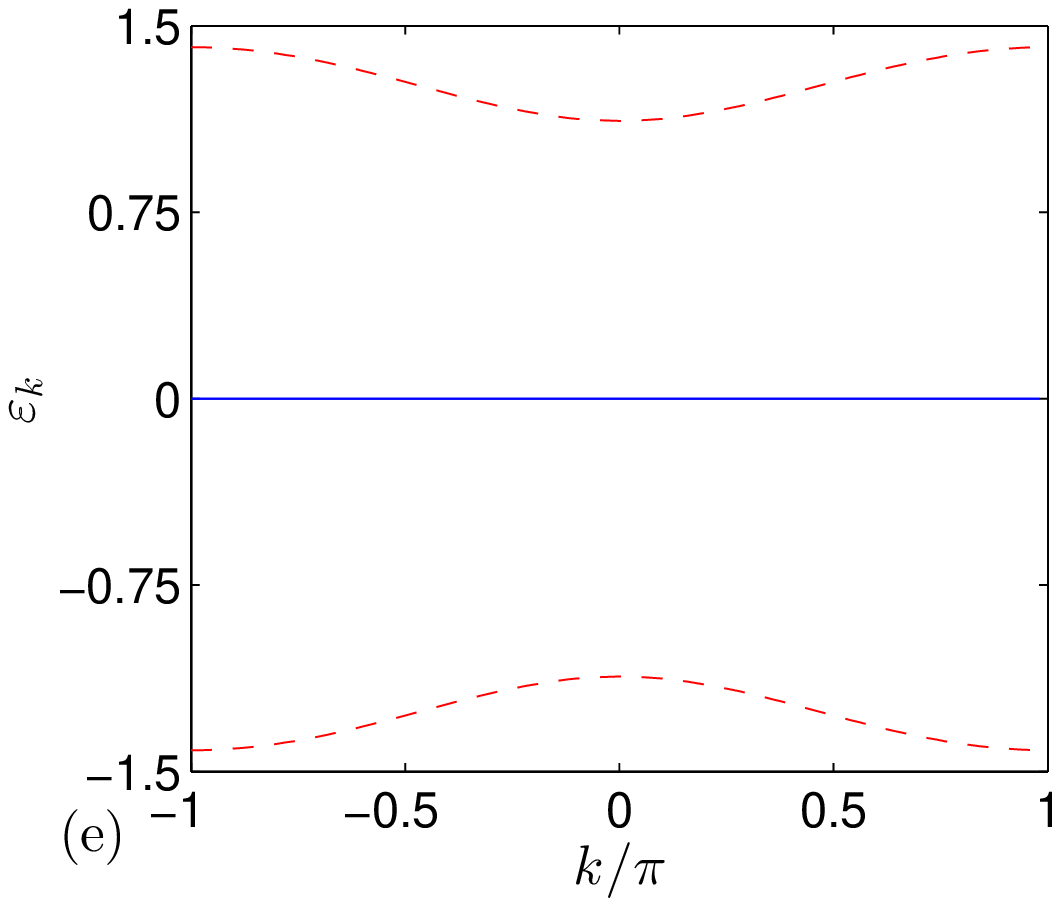} %
\includegraphics[ bb=25 242 340 511, width=0.3\textwidth, clip]{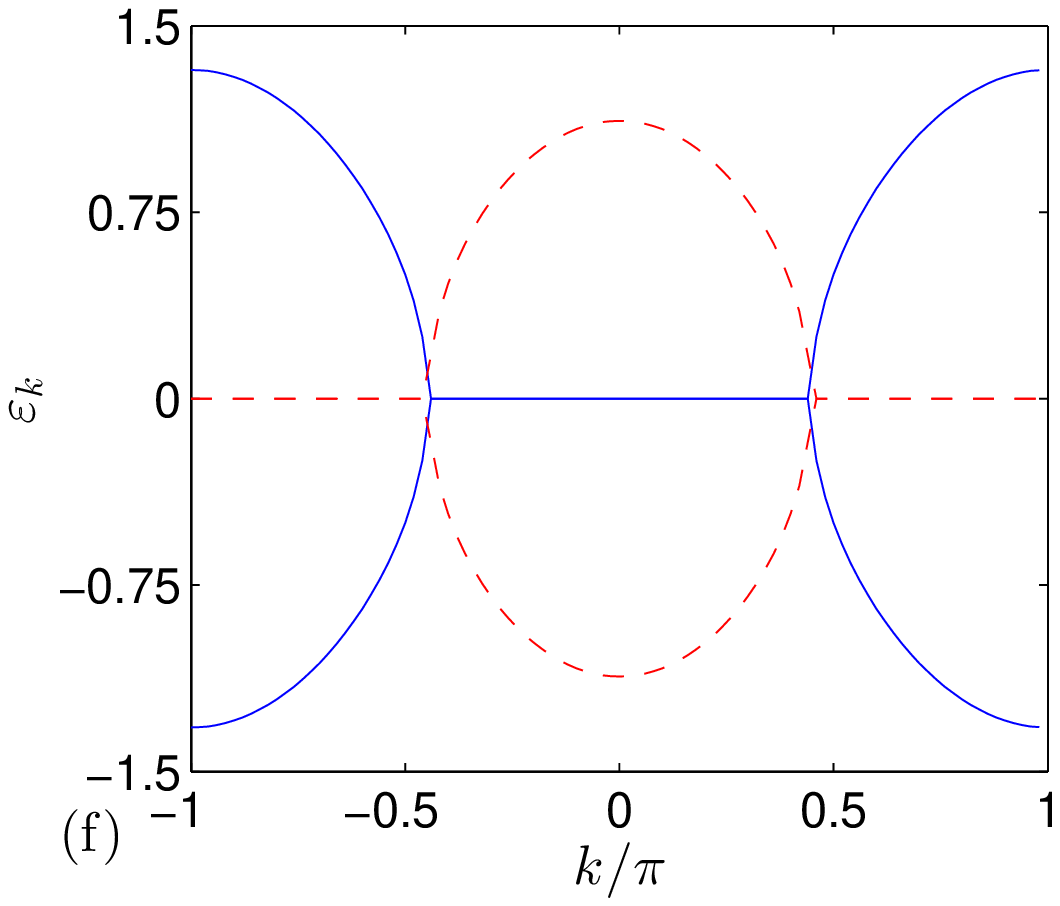}
\caption{(Color online) Plots of the structure of the energy band described
by Eq. (\protect\ref{energy band}) at some points fixed in Fig. \protect\ref%
{fig2}. The blue line means the real part of $\pm \protect\varepsilon _{k}$,
the red dashed line means the imaginary part of $\pm \protect\varepsilon _{k}
$, respectively. $J=1/2$ for all figures. Graphs (a), (c), (d), and (f) show
the energy band situation in the nontrivial area where the EPs exist.
Meanwhile (b) and (e) display the situation in the trivial area where $%
\protect\varepsilon _{k}$ is always non-zero.}
\label{fig4}
\end{figure*}

In fact, it can be seen as a SSH chain system, in which the hopping
amplitudes of the chain is staggered. Fig. \ref{fig1} sketches the geometry
of the system. Following the same step we did in Sec. \ref{Model and phase
diagram}, it is quite easy to find the core matrix of $H=\sum_{\boldsymbol{k}%
}H_{\boldsymbol{k}}$ is
\begin{equation}
h_{k}=-J\left(
\begin{array}{cc}
i\frac{\gamma }{J} & w+ve^{-ik} \\
w+ve^{ik} & -i\frac{\gamma }{J}%
\end{array}%
\right) ,
\end{equation}%
where $w=1-\delta $, $v=1+\delta $ and the wave vector $k=2\pi n/N-\pi $, $%
(n=0,1,...,N-1)$. The phase boundary can be obtained by the zero points of
the spectrum

\begin{eqnarray}
\varepsilon _{k} &=&\pm J\sqrt{4wv\cos ^{2}\left( \frac{k}{2}\right) +\left(
w-v\right) ^{2}-\left( \frac{\gamma }{J}\right) ^{2}}  \notag \\
&=&\pm 2J\sqrt{\left( 1-\delta ^{2}\right) \cos ^{2}\left( \frac{k}{2}%
\right) +\delta ^{2}-\left( \frac{\gamma }{2J}\right) ^{2}},
\label{energy band}
\end{eqnarray}%
with the eigenvector%
\begin{equation}
\left\vert \psi _{k}^{\pm }\right\rangle =\frac{1}{\sqrt{\Delta }}\left(
\begin{array}{c}
\frac{1}{J\left( w+ve^{ik}\right) }\left( i\gamma \mp \varepsilon
_{k}\right)  \\
1%
\end{array}%
\right) ,
\end{equation}%
which are normalized by the Dirac normalization factor%
\begin{equation}
\Delta =\frac{\varepsilon _{k}^{2}+\left\vert \varepsilon _{k}\right\vert
^{2}\mp 2\gamma \mathrm{Im}\varepsilon _{k}+2\gamma ^{2}}{\varepsilon
_{k}^{2}+\gamma ^{2}}.
\end{equation}%
From $\varepsilon _{k}=0$, we have equations%
\begin{equation}
\left( 1-\delta ^{2}\right) \cos ^{2}\left( \frac{k}{2}\right) +\delta
^{2}-\left( \frac{\gamma }{2J}\right) ^{2}=0.
\end{equation}%
It is obvious that the phase diagram of this single-chain model is as same
as what is shown in Fig. \ref{fig2}. In Fig. \ref{fig4}, we provide more
details about the energy band structure of different parameters which are
marked in Fig. \ref{fig2}, the numerical result accords with the analysis
above.

The $2$D vector field $F(k)$\ is still effective and reduces to $F(k)$,
which displays the topological property as the kink for $k$\ in $1$D, as
shown along $k_{x}$\ direction with $k_{y}=\pm \pi /2$\ in Fig. \ref{fig3}%
(a). A pair of kinks will meet with each other at the phase boundary, and
disappear as the parameters reach the unbroken area, which is the same as
the $2$D system result.

\section{Symmetry protection of kinks}

\label{Symmetry protection of kinks}

As we mentioned in the introduction, the kink represented by $F(k)$ in the
single-chain system is a topological invariant, which is protected by the
symmetry of the Hamiltonian. To see this point, we firstly introduce the
eigenstate of the $h_{k}$
\begin{eqnarray}
\left\vert \varphi _{k}^{\pm }\right\rangle &=&\frac{1}{\sqrt{\Omega }}%
\left(
\begin{array}{c}
\frac{1}{J\left( w+ve^{ik}\right) }\left( i\gamma \mp \varepsilon _{k}\right)
\\
1%
\end{array}%
\right) ,  \notag \\
\left\vert \phi _{k}^{\pm }\right\rangle &=&\frac{1}{\sqrt{\Omega ^{\ast }}}%
\left(
\begin{array}{c}
\frac{1}{J\left( w+ve^{ik}\right) }\left( -i\gamma \mp \varepsilon
_{k}^{\ast }\right) \\
1%
\end{array}%
\right) ,
\end{eqnarray}%
which are normalized by the biorthogonal normalization with%
\begin{equation}
\Omega =\frac{2\varepsilon _{k}\left( \varepsilon _{k}+i\gamma \right) }{%
\varepsilon _{k}^{2}+\gamma ^{2}},
\end{equation}%
and satisfy%
\begin{eqnarray}
\left\langle \phi _{k}^{a}\right\vert \varphi _{k}^{b}\rangle &=&\delta
_{a,b},\varepsilon _{k}^{2}>0,  \notag \\
\left\langle \phi _{k}^{a}\right\vert \varphi _{k}^{b}\rangle &=&\delta
_{a,-b},\varepsilon _{k}^{2}<0.
\end{eqnarray}%
Then we define
\begin{equation}
Q_{x}=\sigma _{x}\mathcal{T}.
\end{equation}%
This operator can be seen as a combination of the inversion operator $%
\mathcal{P}$ and time-reversal operator $\mathcal{T}$, and leads to $\left[
Q_{x},h_{k}\right] =0$. So
\begin{equation}
Q_{x}\left\vert \varphi _{k}^{-}\right\rangle =\frac{1}{\sqrt{\Omega ^{\ast }%
}}\frac{-i\gamma +\varepsilon _{k}^{\ast }}{J\left( w+ve^{-ik}\right) }%
\left(
\begin{array}{c}
\frac{1}{J\left( w+ve^{ik}\right) }\left( i\gamma +\varepsilon _{k}^{\ast
}\right) \\
1%
\end{array}%
\right)
\end{equation}%
is still a eigenstate of $h_{k}$. When $\varepsilon _{k}^{2}>0$%
\begin{equation}
Q_{x}\left\vert \varphi _{k}^{-}\right\rangle =\sqrt{\frac{w+ve^{ik}}{%
w+ve^{-ik}}}\left\vert \varphi _{k}^{-}\right\rangle ,
\end{equation}%
when $\varepsilon _{k}^{2}<0$%
\begin{equation}
Q_{x}\left\vert \varphi _{k}^{-}\right\rangle =-\frac{i\gamma +\varepsilon
_{k}}{J\left( w+ve^{-ik}\right) }\left\vert \varphi _{k}^{+}\right\rangle ,
\end{equation}%
the symmetry breaking happens. According to the above conclusion, a
straightforward calculation of $\left\vert \left\langle Q_{x}^{\mathrm{B}%
}\right\rangle _{k}\right\vert =\left\vert \left\langle \phi
_{k}^{-}\right\vert Q_{x}\left\vert \varphi _{k}^{-}\right\rangle
\right\vert $ shows that%
\begin{equation}
\left\vert \left\langle Q_{x}^{\mathrm{B}}\right\rangle _{k}\right\vert
=\left\{
\begin{array}{c}
1,\varepsilon _{k}\text{ is real} \\
0,\varepsilon _{k}\text{ is imagine}%
\end{array}%
\right. ,
\end{equation}%
which indicates the topological property of the EP as a critical point of
values $0$ and $1$ in the basis of the biorthogonal eigenvectors.

\begin{figure}[tbp]
\includegraphics[ bb=43 542 429 754, width=0.45\textwidth, clip]{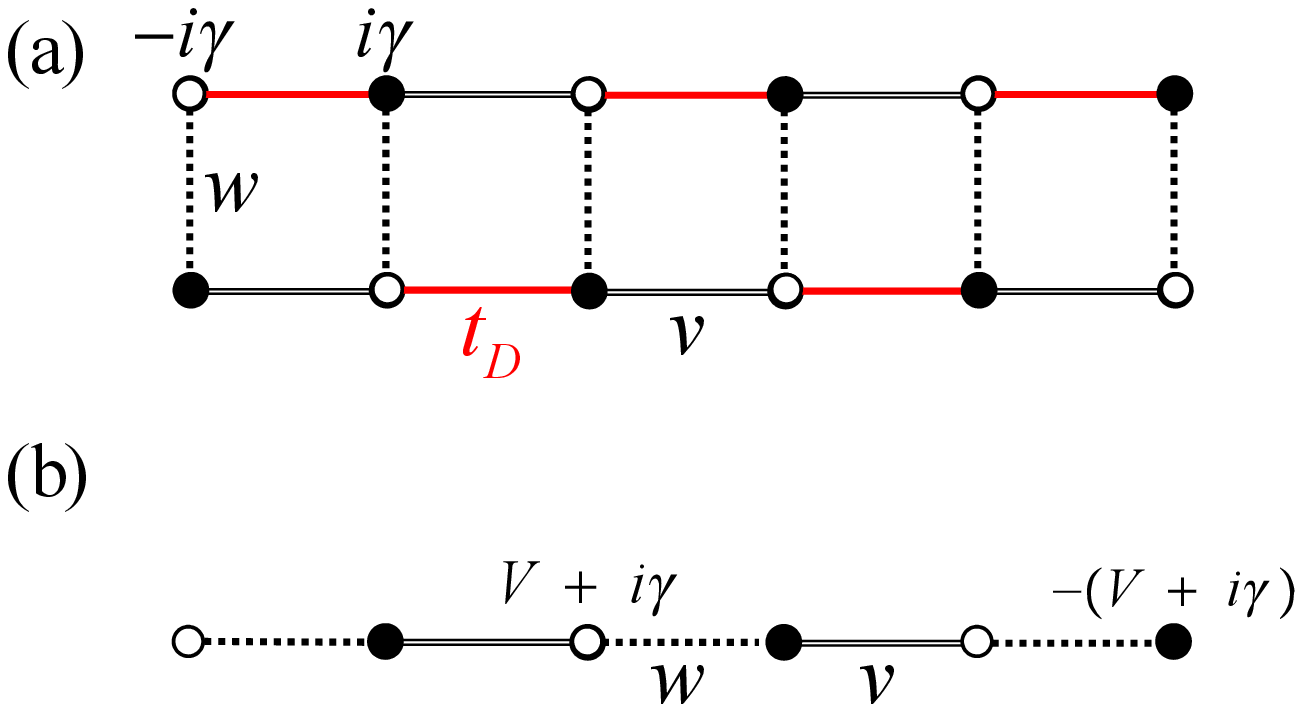}
\caption{(Color online) Schematics of two kinds of perturbation on the SSH
chain at EPs. (a)Extra hopping term (red line) with amplitude $t_{\mathrm{D}%
} $. (b)Staggered on-site real potentials on two sublattices indicated by
filled and empty circles, respectively. In the case of (a), the topology of
EPs is an invariant under the perturbations, while the EPs is eliminated for
non-zero $V$ in the case (b).}
\label{fig5}
\end{figure}

\section{Perturbations}

\label{Perturbations}

In a Hermitian system, the topological invariant for the topological
boundary is essentially the topologically unavoidable band touching points
\cite{Li2,SCZ}. For a non-Hermitian one, as we proved before, it is the
topological unavoidable EPs. Once the parameters of the system change within
some particular areas, the EPs always exist. This topological feature may be
robust for some kinds of perturbation but fragile for others. We consider
two kinds of perturbations, with an extra hopping across the neighboring
site and staggered on-site real potential, respectively. Fig. \ref{fig5}
sketches the structures of these two cases. We will focus on the effects of
the extra terms on the existence of the EPs.

For the first case, the Hamiltonian can be written as%
\begin{equation}
H_{\mathrm{D}}=H-Jt_{\mathrm{D}}\sum_{l=1}^{N}(a_{l}^{\dagger }b_{l+1}+%
\mathrm{H.c.}),
\end{equation}%
where $t_{\mathrm{D}}$\ denotes the extra hopping amplitude. Based on the
Fourier transformations, we still have%
\begin{equation}
H_{\mathrm{D}}=\sum_{k}(a_{k}^{\dagger },b_{k}^{\dagger })h_{k}^{\mathrm{D}%
}\left(
\begin{array}{c}
a_{k} \\
b_{k}%
\end{array}%
\right) ,
\end{equation}%
and%
\begin{equation}
h_{k}^{\mathrm{D}}=h_{k}-Jt_{\mathrm{D}}\left(
\begin{array}{cc}
0 & e^{-ik} \\
e^{ik} & 0%
\end{array}%
\right) .
\end{equation}%
The spectrum is%
\begin{equation}
\varepsilon _{k}^{\mathrm{D}}=\pm J\sqrt{4w\left( v+t_{\mathrm{D}}\right)
\cos ^{2}\left( \frac{k}{2}\right) +\left[ w-\left( v+t_{\mathrm{D}}\right) %
\right] ^{2}-\left( \frac{\gamma }{J}\right) ^{2}},
\end{equation}%
which only has a shift on $v$, i.e., $v\rightarrow v+t_{\mathrm{D}}$, from
the spectrum $\varepsilon _{k}$ in Eq. (\ref{energy band}). Thus the zero
energy point can be obtained directly as following. There are still three
types of zero energy points. (I) $\delta =1$ or $-\left( 1+t_{\mathrm{D}%
}\right) ,\frac{\gamma }{2J}=\pm \left( 1+\frac{t_{\mathrm{D}}}{2}\right) $.
(II) $k_{c}=\pm \pi ,\delta =\pm \frac{\gamma }{2J}-\frac{t_{\mathrm{D}}}{2}$
or $k_{c}=0,\frac{\gamma }{2J}=\pm \left( 1+\frac{t_{\mathrm{D}}}{2}\right) $%
. (III) When $\gamma $ and $\delta $ are in the area%
\begin{eqnarray}
\frac{\gamma }{2J} &\in &\left( -\left\vert 1+\frac{t_{\mathrm{D}}}{2}%
\right\vert ,\left\vert 1+\frac{t_{\mathrm{D}}}{2}\right\vert \right) ,
\notag \\
\delta &\in &\left( -\left\vert \frac{\gamma }{2J}\right\vert -\frac{t_{%
\mathrm{D}}}{2},\left\vert \frac{\gamma }{2J}\right\vert -\frac{t_{\mathrm{D}%
}}{2}\right) ,
\end{eqnarray}%
or%
\begin{eqnarray}
\frac{\gamma }{2J} &\in &\left( -\infty ,-\left\vert 1+\frac{t_{\mathrm{D}}}{%
2}\right\vert \right) \cup \left( \left\vert 1+\frac{t_{\mathrm{D}}}{2}%
\right\vert ,+\infty \right) , \\
\delta &\in &\left( -\infty ,\left\vert \frac{\gamma }{2J}\right\vert -\frac{%
t_{\mathrm{D}}}{2}\right) \cup \left( \left\vert \frac{\gamma }{2J}%
\right\vert -\frac{t_{\mathrm{D}}}{2},+\infty \right) ,  \notag
\end{eqnarray}%
there are always two values of $k$: $\pm k_{c},k_{c}\in \left( 0,\pi \right)
$, which satisfy%
\begin{equation}
\frac{\left( \frac{\gamma }{2J}\right) ^{2}-\left( \delta +\frac{t_{\mathrm{D%
}}}{2}\right) ^{2}}{\left( 1-\delta \right) \left( 1+\delta +t_{\mathrm{D}%
}\right) }=\cos ^{2}\left( \frac{k_{c}}{2}\right) .
\end{equation}%
It is clear that for small $t_{\mathrm{D}}$, the phase diagram changes a
little comparing to the case with zero $t_{\mathrm{D}}$, but keep the
original geometry, which means the Hamiltonian still satisfies $\left[
Q_{x},h_{k}^{\mathrm{D}}\right] =0$. The position of the EP, $k_{c}$, shifts
a little without changing the original topology, i.e., the topological
charge of the kink. Then the EP is a topological invariant under the
perturbation from the $t_{\mathrm{D}}$ term.

For the second case, the Hamiltonian can be written as%
\begin{equation}
H_{\mathrm{V}}=H+V\sum_{j=1}^{N}(b_{j}^{\dagger }b_{j}-a_{j}^{\dagger }a_{j})%
\mathrm{,}
\end{equation}%
which indicates that particles on different sub-lattices have opposite real
chemical potentials. The extra potentials do not break the translational
symmetry.\ By the same procedure, we have%
\begin{equation}
H_{\mathrm{V}}=\sum_{k}(a_{k}^{\dagger },b_{k}^{\dagger })h_{k}^{\mathrm{v}%
}\left(
\begin{array}{c}
a_{k} \\
b_{k}%
\end{array}%
\right) ,
\end{equation}%
and%
\begin{equation}
h_{k}^{\mathrm{v}}=-J\left(
\begin{array}{cc}
\frac{V+i\gamma }{J} & w+ve^{-ik} \\
w+ve^{ik} & -\frac{V+i\gamma }{J}%
\end{array}%
\right) .
\end{equation}%
The spectrum is%
\begin{eqnarray}
\varepsilon _{k}^{\mathrm{v}} &=&\pm J\sqrt{4wv\cos ^{2}\left( \frac{k}{2}%
\right) +\left( w-v\right) ^{2}+\left( \frac{V+i\gamma }{J}\right) ^{2}}
\notag \\
&=&\pm 2J\sqrt{\left( 1-\delta ^{2}\right) \cos ^{2}\left( \frac{k}{2}%
\right) +\delta ^{2}+\left( \frac{V+i\gamma }{2J}\right) ^{2}},
\end{eqnarray}%
which clearly shows that the nonzero $V$\ can let the energy be complex with
$\gamma \neq 0$, i.e., no real energy exist, and destroy the\ topological
EPs. This can be associated with the inversion symmetry $\mathcal{P}$
broken, which leads to $\left[ Q_{x},h_{k}^{\mathrm{V}}\right] \neq 0$,
leave the EP without the symmetry protection.

\section{Summary}

\label{Summary}

In this paper, we study $M$ coupled non-Hermitian Rice-Mele chains with
length $2N$. The EPs appeared in this model can be two isolated points in $2$%
D $\boldsymbol{k}$-plane for infinite $M$ and $N$ or $2$D thermodynamic
limit. It is shown that two isolated EPs are topological defects of an
auxiliary field achieved by mapping the the Bloch states of the model onto a
$2$D real vector field in $\boldsymbol{k}$-plane, with topological charge $%
\pm 1/2$. Furthermore, we extend this analysis to finite $M$ cases with
periodic and open boundary conditions and find that the auxiliary field on
the discrete $\boldsymbol{k}$-space for finite $M$ is the projection of the
infinite one for the $2$D model. It is shown that the EPs in finite $M$
systems still possess topological features. Besides, we focus on the single
chain system and find that two topological defects in $2$D $\boldsymbol{k}$
space reduce to a pair of kinks in $1$D $k$ space. The corresponding
topological invariant is protected by the combined inversion and
time-reversal symmetry. At last, we show the robust topological feature of
the EPs in non-Hermitian systems is as same as the band touching points in
Hermitian systems.

Since it proves that the topological invariants for a quasi-$1$D system can
be extracted from the projection of the corresponding $2$D limit system on
it, and the topological feature for some quasi-$1$D systems is not easy to
describe due to there is only one variable in $\boldsymbol{k}$ space. This
work also provides an alternative way to capture the topological feature in
the quasi-$1$D system by exploring the $2$D expansion of the original system.

\acknowledgments We acknowledge the support of Chinese National Natural Science Foundation (Grant No. 11874225).


\begin{thebibliography}{99}
\bibitem{Bender} C. M. Bender and S. Boettcher, Phys. Rev. Lett. \textbf{80,}
5243 (1998); C. M. Bender, D. C. Brody, H. F. Jones, Phys. Rev. Lett.
\textbf{89,} 270401 (2002); Phys. Rev. Lett. \textbf{98,} 040403 (2007); C.
M. Bender, D. W. Hook, P. N. Meisinger, and Q. H. Wang, Phys. Rev. Lett.
\textbf{104,} 061601 (2010).

\bibitem{Longhi} S. Longhi, Phys. Rev. Lett. \textbf{105,} 013903 (2010);

\bibitem{West} C. T. West, T. Kottos, and T. Prosen, Phys. Rev. Lett.
\textbf{104, }054102 (2010).

\bibitem{LC} C. Li and Z. Song, Phys. Rev. A \textbf{91,} 062104 (2015); C.
Li, L. Jin, and Z. Song, Phys. Rev. A 95, 022125 (2017).

\bibitem{Yang} W. J. Chen, \c{S}. K. \"{O}zdemir, G. M. Zhao, J. Wiersig,
and L. Yang, Nature \textbf{548,} 192-196 (2017).

\bibitem{Zhang} J. Zhang, et al., Nat. Photon. \textbf{12,} 479--484 (2018).

\bibitem{Longhi1} S. Longhi, Opt. Lett. 43, 2929 (2018).

\bibitem{Goldzak} T. Goldzak, A. A. Mailybaev, and N. Moiseyev, Phys. Rev.
Lett. \textbf{120,} 013901 (2018).

\bibitem{Yi} C. H. Yi, J. Kullig, and J. Wiersig, Phys. Rev. Lett. \textbf{%
120,} 093902 (2018).

\bibitem{Heiss} W. D. Heiss, J. Phys. A: Math. Theor. \textbf{45,} 444016
(2012).

\bibitem{Ali1} A. Mostafazadeh, J. Math. Phys. \textbf{43,} 205-214 (2002);
J. Phys. A: Math. Gen. \textbf{36,} 7081-7091 (2003); J. Phys. A: Math. Gen.
\textbf{37,} 11645-11679 (2004).

\bibitem{Ali2} A. Mostafazadeh and H. J. Mehri-Dehnavi, Phys. A: Math.
Theor. \textbf{42,} 125303 (2009); A. Mostafazadeh, Phys. Rev. Lett. \textbf{%
102,} 220402 (2009); Phys. Rev. A \textbf{80,} 032711 (2009); J. Phys. A:
Math. Theor. \textbf{44,} 375302 (2011); A. Mostafazadeh and M. Sarisaman,
Phys. Lett. A \textbf{375,} 3387-3391 (2011).

\bibitem{Esaki} K. Esaki, M. Sato, K. Hasebe, and M. Kohmoto, Phys. Rev. B
\textbf{84,} 205128 (2011).

\bibitem{Tony} T. E. Lee, Phys. Rev. Lett. \textbf{116,} 133903 (2016).

\bibitem{Li1} C. Li, G. Zhang, and Z. Song, Phys. Rev. A \textbf{94,} 052113
(2016); C. Li, X. Z. Zhang, G. Zhang, and Z. Song, Phys. Rev. B \textbf{97,}
115436 (2018).

\bibitem{Lin} S. Lin, G. Zhang, and Z. Song, Sci. Rep 6, 31953 (2016).

\bibitem{Nori} D. Leykam, K. Y. Bliokh, C. L. Huang, Y. D. Chong, and F.
Nori, Phys. Rev. Lett. \textbf{118,} 040401 (2017).

\bibitem{Fu} H. T. Shen, B. Zhen, and L. Fu, Phys. Rev. Lett. \textbf{120,}
146402 (2018).

\bibitem{Yao} S. Y. Yao and Z. Wang, Phys. Rev. Lett. \textbf{121,} 086803
(2018).

\bibitem{Gong} Z. Gong, Y. Ashida, K. Kawabata, K. Takasan, S. Higashikawa,
and M. Ueda, Phys. Rev. X \textbf{8,} 031079 (2018).

\bibitem{Ding} K. Ding, G. C. Ma, M. Xiao, Z. Q. Zhang, and C. T. Chan,
Phys. Rev. X \textbf{6,} 021007 (2016).

\bibitem{Dop} J. Dopple, et al. Nature 537, 76--79 (2016).

\bibitem{Weim} S. Weimann, et al. Nat. Mat. 16, 433--438 (2016).

\bibitem{Feng} B. Midya, H. Zhao, and L. Feng, Nat. Commun. \textbf{9,} 2674
(2018).

\bibitem{Niu} D. Xiao, M. C. Chang, and Q. Niu, Rev. Mod. Phys. \textbf{82,}
1959 (2010)

\bibitem{Kane} M. Z. Hasan and C. L. Kane, Rev. Mod. Phys. \textbf{82,} 3045
(2010).

\bibitem{ZhangSQ} X. L. Qi and S. C. Zhang, Rev. Mod. Phys. \textbf{83,}
1057 (2011).

\bibitem{Burkov} A. A. Burkov and L. Balents, Phys. Rev. Lett. \textbf{107,}
127205 (2011).

\bibitem{Xu} G. Xu, H. Weng, Z. Wang, X. Dai, and Z. Fang, Phys. Rev. Lett.
\textbf{107,} 186806 (2011).

\bibitem{Young} S. M. Young, S. Zaheer, J. C. Y. Teo, C. L. Kane, E. J.
Mele, and A. M. Rappe, Phys. Rev. Lett. \textbf{108,} 140405 (2012).

\bibitem{Sama} K. Sun, W. V. Liu, A. Hemmerich, and D. Sama, Nature. Phys.
\textbf{8,} 67--70 (2012).

\bibitem{Weng} H. Weng, C. Fang, Z. Fang, B. A. Bernevig, and X. Dai, Phys.
Rev. X \textbf{5,} 011029 (2015).

\bibitem{Huang} S. M. Huang, et al. Nat. Commun. \textbf{6,} 7373 (2015).

\bibitem{Hou} J. M. Hou, Phys. Rev. Lett. \textbf{111,} 130403 (2013).

\bibitem{Liu} Z. K. Liu, et al. Science \textbf{343,} 864--867 (2014).

\bibitem{Neupane} M. Neupane, et al. Nat. Commun. \textbf{5,} 3786 (2014).

\bibitem{SXu} S. Y. Xu, et al. Science \textbf{349,} 613--617 (2015).

\bibitem{Lv} B. Q. Lv, et al. Phys. Rev. X \textbf{5,} 031013 (2015).

\bibitem{Lu} L. Lu, et al. Science \textbf{349,} 622--624 (2015).

\bibitem{RM} M. J. Rice and E. J. Mele, Phys. Rev. Lett. \textbf{49}, 1455
(1982).

\bibitem{Lin1} S. Lin, X. Z. Zhang, and Z. Song, Phys, Rev. A \textbf{92},
012117 (2015).

\bibitem{WR} R. Wang, X. Z. Zhang, and Z. Song, Phys. Rev. A \textbf{98},
042120 (2018).

\bibitem{SSH} W. P. Su, J. R. Schrieffer, and A. J. Heeger, Phys. Rev. B
\textbf{22}, 2099 (1980).

\bibitem{As} J. K. Asb\'{o}th, L. Oroszl\'{a}ny, and A. P\'{a}lyi, \textit{A
Short Course on Topological Insulators}, Lecture Notes in Physics 919
(Springer International Publishing, Switzerland, 2016).

\bibitem{Kane1} J. C. Y. Teo and C. L. Kane, Phys. Rev. B \textbf{82},
115120 (2010).

\bibitem{Zhou} Y. Q. Yan and Q. Zhou, Phys. Rev. Lett. \textbf{120}, 235302
(2018).

\bibitem{Li2} C. Li, S. Lin, G. Zhang, and Z. Song, Phys. Rev. B \textbf{96}%
, 125418 (2017).

\bibitem{SCZ} J. Wang and S. C. Zhang, Nat. Mater. \textbf{16}, 1062-1067
(2017).
\end{thebibliography}
\end{document}